# Twitter's Agenda-Setting Role: A Study of Twitter Strategy for Political Diversion

Yuyang Chen[1#]   Xiaoyu Cui[2#]   Yunjie Song[3]   Manli Wu[4*]

[1#] Putnam Science Academy, 18 Maple St, Putnam, CT 06260, USA
   e-mail: Aftersilver001@gmail.com
[2#] Princeton High School, 151 Moore St, Princeton, NJ 08540, USA
   e-mail: raincui1020@gmail.com
[3] Zhengzhou University, No.100 Science Road, Zhengzhou, Henan China 45001
   e-mail: songwork1007@gmail.com
[4*] Huazhong University of Science and Technology, Wuhan China 430022
   e-mail: mlwu@hust.edu.cn

**Abstract**

This study verified the effectiveness of Donald Trump's Twitter campaign in guiding agenda-setting and deflecting political risk and examined Trump's Twitter communication strategy and explores the communication effects of his tweet content during Covid-19 pandemic. We collected all tweets posted by Trump on the Twitter platform from January 1, 2020 to December 31, 2020. We used Ordinary Least Squares (OLS) regression analysis with a fixed effects model to analyze the existence of the Twitter strategy. The correlation between the number of confirmed daily Covid-19 diagnoses and the number of particular thematic tweets was investigated using time series analysis. Empirical analysis revealed Twitter's strategy is used to divert public attention from negative Covid-19 reports during the epidemic, and it posts a powerful political communication effect on Twitter. However, findings suggest that Trump did not use false claims to divert political risk and shape public opinion.

**Key Words:**   Agenda Setting, Political Communication, News Media.

**Introduction**

In 2016 Donald Trump broke the shackles of traditional media and used the instantaneous, interactive and universal nature of Twitter to attack his political opponents and counter negative public opinion, attracting widespread attention and successfully defeating Hillary to win the 45th presidency. This phenomenon has led to research on the content and Twitter strategies of political figures' tweets [1-7]. As Trump often attacked the mainstream media and political opponents on Twitter, spreading controversial statements, much of the existing research has focused on how Twitter helped to explain Trump's victory and gain widespread attention [8-10]. In contrast, there is less research on how Trump diverted public attention. We have observed that Trump may employ diversionary tactics to distract the public when negative news that poses a political threat to Trump. For example, on 19 November 2016, the Trump University lawsuit was finally settled for $25 million.

Still, on the same day, President-elect Donald Trump publicly accused the Hamilton actor of calling on Pence to "uphold our American values" The act was rude and the cast was asked to apologize to Pence. A Google Trends analysis found that the public was more concerned about Broadway than the Trump University incident[11] and this shift appears to have been effective. These phenomena are consistent with what George P. Lakoff [12] has identified as Trump's "Diversion" strategy in his Tweets, which diverts public attention from negative events to unrelated ones. This diversion strategy has attracted our interest, this paper explores the process and effects of this strategy in the context of the COVID-19 epidemic, considering the role of the agenda setting in Trump's Tweets.

In 2020, COVID-19 overlaps with an election year, which means Donald Trump must balance COVID-19epidemic while maintaining his electoral competence. However, with the outbreak of the epidemic worldwide and the gradual increase of negative news such as school closures, unemployment, and economic contraction, voters have become discontent with Trump's response to the epidemic. In this unfavorable situation, Trump used Twitter to downplay the threat of COVID-19 to the US, as well as reinforced his own "achievements" in terms of jobs and the economy during his administration. But as Nancy Pelosi argues, Trump's investigation of China's issues such as the "source of the outbreak" ultimately attempts to "divert" domestic attention away from the fight against the epidemic and from solving pressing problems such as the US economy and jobs diversion" [14]. To investigate the effectiveness of this Twitter strategy in directing agenda settings and diverting political risk, this study analyzed Trump's Twitter diversion strategy and the dissemination of related Tweets content. This study not only helps to understand the deeper motives behind his Twitter behavior, but also to better understand how Twitter has become a tool for political figures to disseminate their political ideas.

**Agenda setting**

The agenda-setting theory points out[15] that "the media has the function of setting public agenda", i.e., influencing public perceptions of issues by emphasizing that different issues have different levels of importance, which in turn becomes the central issue of public opinion discussions. This theory suggests that public opinion can be influenced and even shaped Therefore it is of great theoretical and practical importance to improve the effectiveness of political communication through agenda setting. Due to the continuous innovation of the media, the traditional media represented by newspapers, radio, and television have started to lose their power, and the influence of social media is rapidly gaining ground [16-17]. In the traditional media era, the news media, with its news production process and influence, controlled whether particular content reached its audience, so in the early days it was mainly the media that determined the agenda [1], and politicians had limited power to set public agenda. However, with the rise of social media, the media's monopoly was broken [18-19] and the media's advantage in time and speed was challenged, allowing the public to bypass the media to access news facts [16,20]. The US presidential election is a concentrated example of the function of agenda-setting [20,21-23]. Twitter serves as an important platform and tool for politicians to communicate with voters, influencing public opinion, and even shaping their persona. It allows candidates to express their views without going through the gatekeeping process of the mainstream media, so politicians can directly influence the agenda setting [24,25]. Lee, J., & Xu, W. (2018) also confirmed that Twitter has the potential to shape public agenda by examining the Twitter and voter responses of candidates in the 2016 US presidential election[26].

By sorting out agenda-setting theory, political figures can use Twitter to post Tweets to direct people's attention to a specific agenda, thus diverting public attention.

**Twitter strategy**

Twitter is an American social networking platform that allows users to interact with hot topics on Twitter by posting messages (i.e. Tweets). Studies have shown that social media, as represented by Twitter, has reshaped the way information is disseminated and accelerated the speed of information dissemination [27]. Lakoff (2017) has identified four common strategies used by Trump--'Pre-emptive framing' (Being the first to frame an idea), "Diversion" (Divert attention from real issues), "Trial Balloon" (Test public reaction) and "Deflection" (Attack messenger, change direction). Ross, A. S., & Rivers, D. J. (2018) built on Lakoff's proposed deflection strategy, confirmed that deflection is indeed the dominant strategy in Trump Tweets [32], finding that Trump shakes the media's credibility by attacking the media and using his rhetoric as a source of truth. Lee, J., & Xu, W. (2018) study also found that Trump uses his rhetoric as a source of truth by attacking the mainstream media and blaming Clinton on Twitter, Trump earned more "hit the like button" and retweets during the 2016 US election.

As Twitter has the potential to shape the public agenda [26], it has become a strategy for politicians to deflect political risk by using the Twitter agenda-setting function to distract the public. Lee, J., & Xu, W. (2018) found that Trump was more successful than Clinton in drawing public attention. Lewandowsky and Ecker (2020) provide the first empirical test of the hypothesis that Trump used Twitter to divert public attention from negative political news. The findings show that Trump's posting of irrelevant Tweets (Y) increased after the release of negative news about Russia (X). In turn, Trump's move reduced the media's follow-up coverage of the Russia story, successfully achieving a diversion of follow-up coverage of politically threatening topics in critical media outlets (New York Times and ABC News) [33]. In contrast, Ross, A. S., & Rivers, D. J. (2018) conclude, after research that contradicts this, that Trump used little diversionary tactics in his Tweets. 2020 COVID-19epidemic outbreak, Lang, K., & Li, A. X. (2021) analysis confirms that during the epidemic, Trump's Tweets against China have become a tool to deflect domestic criticism [34].

A review of the literature on Twitter tactics reveals that most of the research has focused on deflection tactics represented by Trump's attacks on the media and political enemies. Little research has been conducted on deflection tactics, and scholars have expressed different views on whether deflection tactics are widespread.

**Methodology**

**Analytical Framework**

In our study, we drew on one of Trump's four commonly used Tweets tactics proposed by George P. Lakoff in 2017, the diversionary strategy of diverting public attention from an event. In terms of research methodology, we were inspired by Lewandowsky and Ecker identified the five most frequently mentioned themes (unrelated to the epidemic) by Trump during the epidemic

In the problematic study of the Twitter diversion strategy, we used the negative impact of the epidemic, and the number of Trump-specific topic Tweets as the independent and dependent variables, respectively. We collected "hit the like button" (Daily Likes Counts) and retweets (Daily Retweets Counts) for each thematic Tweet to explore the correlation between the negative impact of the epidemic and the spread of thematic Tweets. The correlation between the negative impact of the epidemic and the spread of thematic Tweets. In the course of our research, we found that Trump often posts controversial and even false Tweets. To investigate whether these Tweets have the same effect of deflecting political risk and shaping public opinion as the thematic Tweets, we collected the number of Tweets about Trump's false claims (Daily False Claims Counts) for regression. The number of Tweets about Trump's false claims (Daily False Claims Counts) was regressed and analyzed.

**Data Collection**

We collected three types of data from January 1 to December 31, 2020.

(1) Daily COVID-19 confirmed diagnosis data. This data is available on the CDC website1.

(2) All Tweet content posted by Trump (Twitter handle @realDonaldTrump) on the Twitter platform, as well as like and retweet data for each tweet. This part of the data can be obtained through the Kaggle website. To ensure the integrity of the data collected, even Tweets deleted by Trump are still recorded.

(3) The number of Tweets related to false claims posted by Trump on Twitter. This can be obtained from the statistics published by the Washington Post.

**Variables**

(1) Log_positive_increase daily number of confirmed COVID-19 diagnoses.

This variable is to quantify the negative impact of the epidemic because the number of confirmed daily COVID-19 diagnoses. Compared to Lewandowsky and Ecker's use of the number of media news reports, the method used in this study is more objective and precise, and largely reduces measurement errors due to different media positions, reporting preferences, and delayed reporting.

(2) Topical Tweets Counts: The five themes of Tweets were identified, including "foreign policy", "economy and job", "domestic politics", "black lives" and "job". politics", "Black Lives Matter movement" and "presidential election campaign"

(3) Daily Likes Counts is the number of "hit the like button" for a particular theme of Tweets, used to measure the spread of Tweets.

(4) Daily Retweets Counts is the number of retweets for a particular topic of Tweets, used to measure the reach of Tweets.

(5) Daily False Claims Counts is the number of Tweets of Trump's false claims.

**Data processing**

The tool for data analysis is Python.

(1) Tweets theme determination.

We conducted a thematic analysis of Trump's Tweets to identify the Tweets that Trump often uses to divert attention. The criteria is that the economy and job have been a proud and repeated issue during the Trump administration. Before the outbreak of COVID-19 epidemic, the unemployment rate in the US had fallen to 3.6% in December 2019, the lowest level in a decade(Figure 1). Foreign policy issues were a key area of Trump's campaign. The "America First" philosophy and preference have successfully packaged him as a leader who defends the national interests of the US and has gained the support of the lower and middle classes. Blaming China for the epidemic is another way of distracting the public. Mover, the issue of domestic politics, represented by immigration, is one of the most obvious issues that oppose the views of Democratic and Republican voters. In the second half of 2020, the shooting of black people by police officers in the US led to protests by black people around the world. "Black Lives Matter movement" and the "presidential election campaign" were added as key words for the second half of 2020.

To further support the rationality of the chosen themes, the top ten agendas presented in the Trump Campaign Announces were analyzed. The keyword--Fighting for You! are jobs, eradicate COVID-19 and ending our reliance on China are related to this them--"economy and job" and "domestic politics". In addition, it is found that the themes chosen in the study overlap with the keywords---jobs, immigration, and China derived by Lewandowsky and Ecker in 2016 election based on Tweets word cloud data and campaign website data. Immigration and China also overlap, and the plausibility of the selected themes is further corroborated. To ensure the robustness of the findings, we additionally added the covid theme. The transfer strategy was considered effective when the number of daily confirmations increased.

(2) Construction of a thematic keyword database.

After the initial data processing, the keywords that appeared many times in the subject-specific Tweets were manually filtered to obtain the final word database. The keywords for covid, foreign policy, economy and job, domestic politics, Black Lives Matter movement, and presidential election campaign were 82, 96, 50, 170, 212, and 212 respectively. The keywords for covid, foreign policy, economy and job, domestic politics, Black Lives Matter movement, and presidential election campaign were 82, 96, 50, 170, 212, and 1269 respectively.

**Models**

The regressions in the empirical evidence all use OLS REGRESSION MODEL, and the program runs and the output of the results are done in python. Considering that the black movement, which only took off in the second half of 2020, and the rising heat of the issue of the upcoming election, the regressions of the first half of the year and the second half of the year were conducted respectively, due to the addition of the black movement and the election theme in the second half of the year, based on the five major themes in the first half of the year.

We use model (1) to study the existence and effect of Trump's Twitter diversion strategy.

$$topic\ tweets_t = \alpha + \beta \log(covid\_increase_t) + \sum_t \mu_t week_t + \varepsilon_t \quad (1)$$

The dependent variable 〖topic tweets〗_t represents the number of relevant topic tweets posted by Trump on day t, and the independent variable log(〖covid_increase〗_t) represents the logarithm of the number of confirmed COVID-19 diagnoses on day t. To ensure the scientific validity of the results and to test whether Trump was avoiding COVID-19-related themes, we added regressions on covid theme Tweets. Finally, because the COVID-19 epidemic is highly propagative and changes significantly in the short term, weekly fixed effects were used to improve the fit of the model.

We change the dependent variable in model (1) by replacing it with the number of "hit the like button", the number of retweets, and the number of false claims, respectively, to test the propagation effect of Trump's Tweets.

**Rusults**

**Descriptive statistics**

Tables 1-2 show the descriptive summary of the variables, the positive case of the first half of the year is lower than that in the second half of the year. The data indicated that the negative impact of Covid-19 pandemic increased in US.

**Table 1. Statistical Summary in the First Half of the Year**

| first year | N | Mean | Std | Min | Max |
| --- | --- | --- | --- | --- | --- |
| positive_increase | 170 | 15429.680 | 13576.410 | 0 | 47032 |
| log_positive_increase | 170 | 6.832 | 4.397 | 0 | 11 |
| covid | 170 | 7.600 | 6.917 | 0 | 49 |
| foreign policy | 170 | 7.824 | 6.036 | 0 | 48 |
| economy and job | 170 | 5.506 | 4.571 | 0 | 28 |
| domestic politics | 170 | 11.829 | 9.236 | 0 | 75 |

The results in Table 2 show that the spread of the COVID-19 epidemic increased in the second half of the year, suggesting that the negative impact of COVID-19 is increasing and that the effectiveness of diversion strategies will be challenged.

**Table 2. Statistical Summary in the Second Half of the Year**

| second half of the year | N | Mean | Std | Min | Max |
| --- | --- | --- | --- | --- | --- |
| positive_increase | 184 | 92728.880 | 63254.830 | 22310 | 242970 |
| log_positive_increase | 184 | 11.230 | 0.629 | 10 | 12 |
| covid | 184 | 5.272 | 4.638 | 0 | 25 |
| foreign policy | 184 | 6.391 | 5.023 | 0 | 26 |
| economy and job | 184 | 6.750 | 5.384 | 0 | 25 |
| domestic politics | 184 | 12.489 | 9.775 | 0 | 57 |
| black lives matter movement | 184 | 8.973 | 7.482 | 0 | 54 |
| presidential election campaign | 184 | 22.293 | 17.107 | 0 | 98 |

### Regression analysis of Twitter strategy

**Table 3. The Regression with the Dependent Variable as Topic Tweets Counts**

| | | log(covid_increase_t) | covid_increase_t | Week FE | Observations | $R^2$ | Adj.$R^2$ |
|---|---|---|---|---|---|---|---|
| (1) | covid | 1.351 | (1.596) | YES | 170 | 0.245 | 0.114 |
| (2) | foreign | 2.666* | (1.420) | YES | 170 | 0.216 | 0.080 |
| (3) | job | 3.224*** | (1.065) | YES | 170 | 0.230 | 0.096 |
| (4) | domestic | 6.434*** | (2.041) | YES | 170 | 0.307 | 0.187 |
| (5) | covid | -14.919 | (11.999) | YES | 184 | 0.314 | 0.195 |
| (6) | foreign | -2.950 | (2.615) | YES | 184 | 0.259 | 0.131 |
| (7) | job | -0.949 | (2.830) | YES | 184 | 0.244 | 0.113 |
| (8) | domestic | -5.739 | (5.104) | YES | 184 | 0.254 | 0.125 |
| (9) | black | -1.444 | (3.921) | YES | 184 | 0.249 | 0.119 |
| (10) | election | -10.991 | (8.953) | YES | 184 | 0.251 | 0.121 |

Note: * significant at the 10% level; ** significant at the 5% level; *** significant at the 1% level, same below.

Table 3 shows the regression results for the OLS regression model, where columns (1)-(4) and (5)-(10) lead the regression results for the first and second half of the year, respectively, with standard deviations in parentheses. If Trump deflects political risk and avoids COVID-19-related reports with the help of theme Tweets, then the model's regression coefficients on specific themes should be positive and statistically significant. Meanwhile, the regression coefficient on covid is insignificant. The results show that the transfer strategy proves to be significant only in the first half of the year. Specifically, the number of epidemic-related topic Tweets did not increase significantly in the first half of 2020 as the number of daily confirmations increased, while the number of topic-specific Tweets used to divert public attention increased accordingly. Of these, the coefficients of foreign policy, economy and job, and domestic politics theme Tweets passed the significance level test of 10%, 1%, and 1% respectively. In terms of numbers, for every 1% increase in the number of confirmed COVID-19 cases on that day, the number of Tweets on these three topics would increase by 2.7, 3.2, and 6.4 respectively, confirming the existence of political diversion strategies on Twitter.

### Regression analysis of communication effects

**Table 4. The Regression with the Dependent Variable as Daily Retweets Counts**

| | | log(covid_increase_t) | covid_increase_t | Week FE | Observations | $R^2$ | Adj.$R^2$ |
|---|---|---|---|---|---|---|---|
| (1) | covid | 1.710 | (2.571) | YES | 170 | 0.522 | 0.439 |

|      |          | log(covid_increase_t) | covid_increase_t | Week FE | Observations | R² | Adj.R² |
|------|----------|-----------------------|------------------|---------|--------------|------|--------|
| (2)  | foreign  | 1.636                 | (2.839)          | YES     | 170          | 0.474 | 0.382 |
| (3)  | job      | 3.120*                | (1.783)          | YES     | 170          | 0.521 | 0.438 |
| (4)  | domestic | 7.175**               | (3.368)          | YES     | 170          | 0.606 | 0.538 |
| (5)  | covid    | -6.728                | (9.254)          | YES     | 184          | 0.198 | 0.059 |
| (6)  | foreign  | -22.802***            | (9.156)          | YES     | 217          | 0.271 | 0.149 |
| (7)  | job      | -8.2000               | (10.784)         | YES     | 184          | 0.183 | 0.041 |
| (8)  | domestic | -8.249                | (12.808)         | YES     | 184          | 0.199 | 0.060 |
| (9)  | black    | -33.485***            | (15.984)         | YES     | 184          | 0.244 | 0.113 |
| (10) | election | -51.468**             | (25.860)         | YES     | 184          | 0.320 | 0.202 |

Note: * significant at the 10% level; ** significant at the 5% level; *** significant at the 1% level, same below.

**Table 5. The Regression with the Dependent Variable as Daily Likes Counts**

|      |          | log(covid_increase_t) | covid_increase_t | Week FE | Observations | R² | Adj.R² |
|------|----------|-----------------------|------------------|---------|--------------|------|--------|
| (1)  | covid    | 0.990                 | (36.084)         | YES     | 184          | 0.252 | 0.122 |
| (2)  | foreign  | -10.688               | (9.484)          | YES     | 170          | 0.364 | 0.254 |
| (3)  | job      | 0.206                 | (7.293)          | YES     | 170          | 0.403 | 0.300 |
| (4)  | domestic | -3.144                | (10.797)         | YES     | 170          | 0.482 | 0.392 |
| (5)  | covid    | 0.990                 | (36.084)         | YES     | 184          | 0.252 | 0.122 |
| (6)  | foreign  | -51.951               | (41.173)         | YES     | 184          | 0.246 | 0.115 |
| (7)  | job      | -32.472               | (41.685)         | YES     | 184          | 0.260 | 0.131 |
| (8)  | domestic | -96.974*              | (56.635)         | YES     | 184          | 0.296 | 0.175 |
| (9)  | black    | -2.092                | (42.272)         | YES     | 184          | 0.303 | 0.182 |
| (10) | election | -102.129              | (80.046)         | YES     | 184          | 0.490 | 0.402 |

Note: * significant at the 10% level; ** significant at the 5% level; *** significant at the 1% level, same below.

**Table 6. The Regression with the Dependent Variable as Daily False Claims Counts**

|      |          | log(covid_increase_t) | covid_increase_t | Week FE | Observations | R² | Adj.R² |
|------|----------|-----------------------|------------------|---------|--------------|------|--------|
| (1)  | covid    | -1.034                | (1.659)          | YES     | 184          | 0.115 | -0.039 |
| (2)  | foreign  | -1.709                | (1.820)          | YES     | 170          | 0.079 | -0.081 |
| (3)  | job      | -0.474                | (1.981)          | YES     | 170          | 0.024 | -0.024 |
| (4)  | domestic | -0.637                | (3.367)          | YES     | 170          | 0.105 | -0.050 |
| (5)  | covid    | 5.071                 | (6.670)          | YES     | 184          | 0.622 | 0.557 |
| (6)  | foreign  | 5.471                 | (7.919)          | YES     | 184          | 0.583 | 0.511 |
| (7)  | job      | 8.789                 | (10.153)         | YES     | 184          | 0.652 | 0.592 |
| (8)  | domestic | 12.011                | (14.731)         | YES     | 184          | 0.627 | 0.563 |
| (9)  | black    | 15.967                | (10.266)         | YES     | 184          | 0.581 | 0.508 |
| (10) | election | 28.941                | (23.052)         | YES     | 184          | 0.621 | 0.555 |

Note: * significant at the 10% level; ** significant at the 5% level; *** significant at the 1% level, same below.

Tables 4-6 show the regression results with the number of retweets, the amount of "hit the like button", and the number of incorrect statements as dependent variables, respectively. Columns (3) and (4) of Table 4 show that the number of daily COVID-19 confirmations in the first half of the year is significantly and positively correlated with the number of Trump retweets on economy and job and domestic politics Tweets, with 10% and 5% significance level, respectively. The number of retweets on economy and job and domestic politics increased by 3.1 and 7.1 respectively for each 1% increase in daily COVID-19 confirmations, indicating that this type of Tweet gained more exposure and dissemination during the epidemic.

There was no strong correlation between the number of diagnoses of COVID-19 and the number of "hit the like button" and false statements. Columns (6) (9) and (10) of Table 4 show that as the number of daily COVID-19 confirmations increased in the second half of the year, the number of retweets on foreign policy, domestic politics and presidential election campaign topics decreased significantly with 5% significance test level. Similarly, the number of Trump's "hit the like button" on domestic politic also decreased significantly in the second half of the year, with 10% significance level. The number of Trump's 'hit the like button on domestic politics also declined significantly in the second half of the year, with a 10% significance level.

**Discussion**

Our empirical study confirms the existence of a diversionary strategy in the 2020 US election, where Trump successfully used Twitter to steer the agenda-setting in the first half of the year, diverting public attention from the epidemic to a certain extent.

Inspired by Lewandowsky and Ecker's study on Twitter diversionary tactics, we draw on their research ideas and optimize the measurement of variables based on them. The original paper quantified the negative impact of political threats in terms of the number of stories about negative news in the New York Times and ABC, however, there is no guarantee that the selected newspapers may have a reporting preference in their neutral position, e.g. the New York Times is biased towards the Democratic Party, and countering Trump's (Republican) allegations that he spreads false news dominates a large proportion of the newspaper's content, which would lead to measurement error. In our study, the negative news came from the threat posed by COVID-19, and we used the number of confirmed COVID-19 diagnoses to quantify the negative impact of COVID-19 more precisely, taking into account the feasibility of data collection while ensuring the objectivity of the data results.

Our study fills a research gap on agenda setting in the context of the epidemic. The findings show that in the face of the sudden COVID-19 epidemic, social media can still influence agendas settings to bypass mainstream media to influence agenda settings and spread political ideas. However, in terms of the timeliness of the agenda set up, the Twitter strategy to influence the agenda was only effective in the short to medium term (first half of the year).

Although the data on false claims in this study is insignificant, we also note that the popularity of Twitter has brought about the prevalence of disinformation [35,36] and has to some extent, exacerbated the polarization of American politics [37]. The epidemic-induced recession has exacerbated the class divide in the US, and social media (Twitter) as social media (Twitter) as a platform for information dissemination has accelerated the spread of political ideas to the lower and middle classes and intensified the process of political polarization in the US. This process has been fueled by partisan conspiracy theories by US politicians [38] and has led to the prevalence of disinformation.

According to data from the Washington Post website[1] (), 227 false claims about coronavirus were posted on Trump's Twitter in 2020, exaggerating the level of epidemic detection and epidemic stigmatization [39-42]. Trump appeased the public with statements that violated the facts, slammed the rival parties, and shirked responsibility. The truth gradually gave way to positions and partisan interests, laying the groundwork for the epidemic outbreak. Therefore, this study has implications for regulating the use of social media and enhancing its credibility in the post-truth era.

Despite confirming the existence of political risk transfer strategies, our research is undeniably limited by the following points. Firstly, unregistered users can view Tweets while cannot "hit the like button" and retweet them. The retweet and like data may be underestimated at the time of data collection and the actual spread of Tweets may be wider. However, this does not affect the significance of our results and conclusions and may even make the actual regression coefficients more significant. Secondly, we cannot tell whether the thematic Tweets posted by Trump were intended as a 'political diversion'. Still, thereis no doubt that in the first half of 2020, the public's attention was diverted to specific thematic Tweets, thus reducing the focus on Trump's poor performance during the epidemic. As far as the regression results are concerned, this does have the effect of shifting political risk. Finally, as the construction of the thematic keyword database is determined by manual screening, omissions are inevitable, leading to some errors between the actual data and the statistics.

Here we propose directions for future research, which can be combined with different periods to study Trump's Twitter strategy in a targeted manner at different times. In addition, existing research focuses more on how politicians use diversionary tactics to distract public from negative news. The generality of this diversionary strategy still needs to be further investigated.

## Data Source

1.https://data.cdc.gov/Case-Surveillance/United-States-COVID-19-Cases-and-Deaths-by-State-o/9mfq-cb36/data
2.https://www.kaggle.com/datasets/headsortails/trump-Twitter-archive
3.https://www.washingtonpost.com/graphics/politics/trump-claims-database/?itid=lk_interstitial_manual_9
4.https://fred.stlouisfed.org.
5.Trump Campaign Announces President Trump's 2nd Term Agenda: Fighting for You! - Trump Nation News

---

[1] https://www.washingtonpost.com/graphics/politics/trump-claims-database/?itid=lk_interstitial_manual_9

# Author information

## Author Contributions

All authors participated in the work. Yuyang Chen and Yunjie Song wrote the initial draft of the manuscript. Yuyang Chen and Xiaoyu Cui collected Data and processed the Data. All authors reviewed and edited the manuscript.

## Corresponding author

Correspondence to Manli Wu, PhD, School of Journalism and Information Communication, Huazhong University of Science and Technology

# Ethics declarations

## Competing interests

There is no competing interests.